\documentclass[runningheads]{llncs}

\usepackage{color}
\usepackage[utf8]{inputenc}
\usepackage{fancyvrb}
\usepackage[usenames,dvipsnames]{xcolor}
\usepackage{listings}
\usepackage{booktabs}
\usepackage[most]{tcolorbox}

\lstdefinelanguage{Alloy}{
  keywords={all, and, as, assert, but, check, disj, else, exactly, extends, fact, for, fun, iden, if, iff, implies, in, Int, let, lone, module, no, none, not, one, open, or, pred, run, set, sig, some, sum, univ},
  keywordstyle=\color{blue}\bfseries,
  comment=[l][\color{Green}\bfseries]{//},
  morecomment=[s][\color{Green}]{/*}{*/},
  stringstyle=\color{red},
  morestring=[b]",
  sensitive=true
}

\lstdefinestyle{AlloyTable}{
  language=Alloy,
  basicstyle=\fontsize{8}{10}\ttfamily,
  numbers=none,
  backgroundcolor=\color{white},
  showspaces=false,
  showstringspaces=false,
  showtabs=false,
  frame=none,
  tabsize=2,
  breaklines=true,
  breakatwhitespace=false,
  aboveskip=0pt,
  belowskip=0pt
}

\tcbset{
    mytextbox/.style={
        colback=gray!10,
        colframe=white,
        boxrule=0pt,
        sharp corners,
        left=2mm,
        right=2mm,
        top=2mm,
        bottom=2mm,
    }
}

\fvset{commandchars=\\\{\}}

\newcommand{\BLUE}[1]{\textcolor{blue}{\textbf{#1}}}
\newcommand{\GREEN}[1]{\textcolor{Green}{\textbf{#1}}}

\usepackage{amssymb}
\usepackage{pifont}
\newcommand{\cmark}{\ding{51}}
\newcommand{\xmark}{\ding{55}}

\usepackage[T1]{fontenc}

\usepackage{graphicx}
\newcommand{\Comment}[1]{}
\newcommand{\CodeIn}[1]{\begin{small}\texttt{#1}\end{small}}
\newenvironment{CodeOut}{\vspace*{0ex}\begin{small}}
                        {\end{small}\vspace*{0ex}}

\newcommand{\Intro}[1]{\emph{#1}}

\newcommand{\ChatGPTUsed}{OpenAI o3-mini}
\newcommand{\DeepSeekUsed}{DeepSeek R1}
\newcommand{\NumSubjects}{11}
\newcommand{\NumTotalSynthesis}{22}
\newcommand{\NumTotalSketching}{11}
\newcommand{\NumTotalTasks}{33}

\begin{document}

\title{On the Effectiveness of Large Language Models in Writing Alloy
  Formulas}

\author{Yang Hong \and
Shan Jiang \and
Yulei Fu \and
Sarfraz Khurshid}

\authorrunning{Y. Hong, S. Jiang, Y. Fu, and S. Khurshid}

\institute{University of Texas at Austin, Austin TX 78712, USA
\email{\{yang22,shanjiang,yuleifu,khurshid\}@utexas.edu}}

\maketitle              

\vspace*{-2ex}
\begin{abstract}

Declarative specifications have a vital role to play in developing
safe and dependable software systems.  Writing specifications
correctly, however, remains particularly challenging.  This paper
presents a controlled experiment on using large language models (LLMs)
to write declarative formulas in the well-known language Alloy.  Our
use of LLMs is three-fold.  One, we employ LLMs to write complete
Alloy formulas from given natural language descriptions (in English).
Two, we employ LLMs to create alternative but equivalent formulas in
Alloy with respect to given Alloy formulas.  Three, we employ LLMs to
complete sketches of Alloy formulas and populate the holes in the
sketches by synthesizing Alloy expressions and operators so that the
completed formulas accurately represent the desired properties (that
are given in natural language).  We conduct the experimental
evaluation using \NumSubjects{} well-studied subject specifications
and employ two popular LLMs, namely ChatGPT and DeepSeek.  The
experimental results show that the LLMs generally perform well in
synthesizing complete Alloy formulas from input properties given in
natural language or in Alloy, and are able to enumerate multiple
unique solutions.  Moreover, the LLMs are also successful at
completing given sketches of Alloy formulas with respect to natural
language descriptions of desired properties (without requiring test
cases).  We believe LLMs offer a very exciting advance in our ability
to write specifications, and can help make specifications take a
pivotal role in software development and enhance our ability to build
robust software.


\end{abstract}


\section{Introduction}
\label{sec:intro}

Declarative languages have long offered immense value in automating
design and implementation of safe and reliable software
systems~\cite{Nugues2006,Ceri1989,Leavens1998,JacksonAlloy2002,Jones1990,Guttag2012}.
However, their true value has not yet been realized.  A key issue is
that their use often necessitates the need for developers to learn
unfamiliar notations that have a very different semantics from
commonly used imperative languages that they use daily to write their
code in.

A major breakthrough in how we write code has occurred in the last few
years with the coming-of-age of large language models
(LLMs)~\cite{ZhaoETALLLMSurvey2024}, such as
ChatGPT\footnote{\url{https://chatgpt.com/}} and
DeepSeek\footnote{\url{https://chat.deepseek.com/}}, which have
revolutionized software development and are now in ubiquitous use.
While these LLMs have massively disrupted traditional software
development, it is yet unclear how much utility they offer in
developing declarative specifications that are often written in
languages that have a very different semantic basis, offer a higher
level of abstraction, and have much less written code available than
commonly used programming languages.

Our focus in this paper is Alloy~\cite{JacksonAlloy2002}, a well-known
declarative language that is based on relational first-order logic and
includes transitive closure.  Specifications written in Alloy can be
analyzed using the Alloy toolset's SAT-based backend that performs
fully automatic bounded exhaustive analysis.  Specifically, the Alloy
analyzer translates formulas in Alloy to propositional logic with
respect to a given bound on the size of the universe of discourse, and
employs state-of-the-art SAT solvers to solve them.  The Alloy
analyzer offers two kinds of analyses.  One, the users can ask it to
create models -- that are termed \Intro{instances} -- of a desired
property, so that the constraint solving problem encodes the given
property and any solution found forms an Alloy instance.  Two, the
users can ask the analyzer to search for a counterexample to a
property, so that the constraint solving problem encodes the negation
to the given property and any solution found serves as a
counterexample.

The Alloy toolset has been used in many applications over the last two
decades, including software design~\cite{CD2Alloy,CDDiff}, test case
generation~\cite{MarinovKhurshidASE2001}, deep analysis of
code~\cite{JacksonVaziriISSTA2000}, repair of faulty
programs~\cite{GopinathETALTACAS2011}, specification-driven
execution~\cite{Samimi2010,Zaeem2010,Squander},
security~\cite{Maldonado-Lopez2014,Margrave}, and
networking~\cite{OpenflowAlloy}.  While Alloy has been deployed in
many contexts, its use remains largely confined to academia.  In our
experience of using and teaching Alloy, a key issue that prevents its
much wider adoption is the learning curve that it poses to developers
who work with imperative languages.

Our insight is that LLMs have the potential to be of immense use in
writing declarative specifications, thereby substantially reducing the
cost of using the Alloy toolset, and more generally, other lightweight
formal methods.  This paper presents a controlled experiment on using
LLMs for creating Alloy specifications.  Our use of LLMs is
three-fold.  One, we employ LLMs to write complete Alloy formulas from
given natural language descriptions (in English).  Two, we employ LLMs
to create alternative but equivalent formulas in Alloy with respect to
given Alloy formulas.  Three, we employ LLMs to complete
sketches~\cite{SolarLazemaPhD2008,WangETALABZ2018ASketch} of Alloy
formulas and populate the holes in the sketches by synthesizing Alloy
expressions and operators so that the completed formulas accurately
represent the desired properties (that are given in natural language).

Alloy's design focuses on analyzability supported by its fully automatic
SAT-based backend makes Alloy an ideal specification language to be
supported by LLMs.  Their outputs can immediately be validated by the
Alloy backend.  For example, an output formula that has no instance
(i.e., is unsatisfiable) is (most likely) invalid.  As another
example, when the input query to the LLM has a reference formula, the
LLM's output can directly be checked for equivalence against the
reference by looking for a counterexample to the asserted equivalence.

We conduct the experimental evaluation using synthesis and sketching
tasks derived from \NumSubjects{} well-studied subject properties
defined over graphs and binary relations, and employ two popular LLMs,
namely ChatGPT (specifically \ChatGPTUsed) and DeepSeek (specifically
\DeepSeekUsed).  A key aspect of our evaluation is that for each
synthesis problem (from English to Alloy and from Alloy to Alloy), we
ask the LLMs to generate multiple equivalent but non-identical Alloy
formulas as solutions, thereby performing a deeper study of how well
the LLMs handle the semantic and syntactic intricacies of the Alloy
language.

The experimental results show that the LLMs generally perform
surprisingly well on synthesizing complete Alloy formulas from input
properties given in natural language or given in Alloy, and are able
to enumerate multiple unique solutions for each property.  Moreover,
the LLMs are also successful at completing given sketches of Alloy
formulas with respect to natural language descriptions of desired
properties (without requiring test cases, which are required by
traditional sketching
techniques~\cite{SolarLazemaPhD2008,WangETALABZ2018ASketch}).

We find the performance of LLMs surprising for three reasons.  One,
Alloy, despite its visibility in academia, is not a mainstream
language; sketching is used even less.  Therefore, only minimal
training data is available.  Moreover, no fine-tuning (Alloy-specific
or otherwise) of the LLMs is performed by us.  In fact, our queries to
the LLMs do not even describe what a sketch is.  We directly give the
synthesis and sketching problems to publicly available LLMs to solve.
Two, Alloy's notation, which allows for succinct formulation of
complex relational properties, makes the language deceptively hard to
master.  Three, creating multiple equivalent formulations of logical
properties requires an in-depth knowledge of logic.  For some of our
subject problems, the LLMs create many correct unique solutions, some
of which have very different logical structures (e.g., quantification
with one variable versus quantification with two variables).  While it
is conceivable that the LLMs simply memorized solutions to some (or
even all) of our queries because they ``saw'' similar ones in their
training data due to the very basic nature of properties of graphs and
binary relations, the ability to generate many (in some cases, over a
dozen) equivalent solutions -- some of which even surprised us and
provided us new insights into how to formalize the properties -- is
noteworthy.

This paper makes the following contributions:

\begin{itemize}
\item {\bf LLM-based synthesis and sketching of Alloy formulas}.  We
  take a three-fold approach for employing LLMs in writing Alloy
  specifications: 1) using natural language properties; 2) using
  reference Alloy formulas; 3) using Alloy sketches.
\item {\bf Evaluation}.  We present an experimental evaluation using
  \NumTotalSynthesis~synthesis tasks and \NumTotalSketching~sketching
  tasks derived from \NumSubjects~well-studied basic properties of
  graphs and binary relations, and two popular LLMs (\ChatGPTUsed{}
  and \DeepSeekUsed).  The experimental results demonstrate the
  effectiveness of using LLMs in writing Alloy formulas.
\end{itemize}
We believe LLMs offer an exciting advance in our ability to utilize
specifications, and the time has come for declarative languages, in
particular, and (lightweight) formal methods, in general, to move to
the forefront of modern tools for building robust and dependable
software systems.


\section{Illustrative overview}
\label{sec:example}

\begin{figure}[!t]
\begin{center}
\begin{tabular}{@{}c@{}|@{\hspace{1ex}}c@{}|@{\hspace{1ex}}c@{}|@{\hspace{1ex}}c@{}}
\begin{minipage}[t]{.2\textwidth}
\raisebox{-\height}{\includegraphics[width=0.3in]{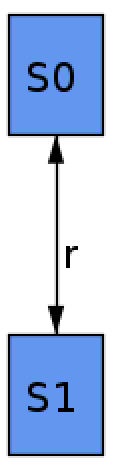}}
\end{minipage}
&
\begin{minipage}[t]{.25\textwidth}
\raisebox{-\height}{\includegraphics[width=1.1in]{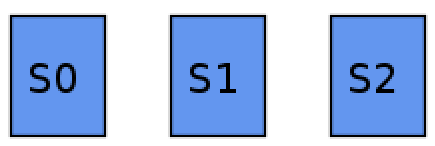}}
\end{minipage}
&
\begin{minipage}[t]{.2\textwidth}
\raisebox{-\height}{\includegraphics[width=0.55in]{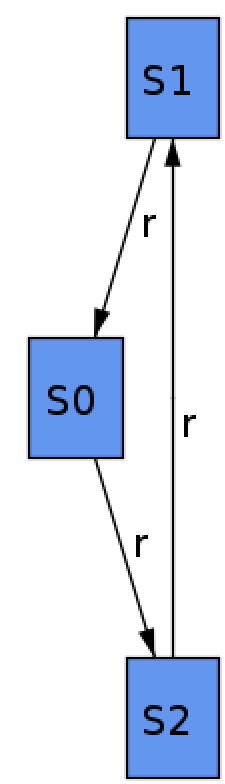}}
\end{minipage}
&
\begin{minipage}[t]{.2\textwidth}
\raisebox{-\height}{\includegraphics[width=0.8in]{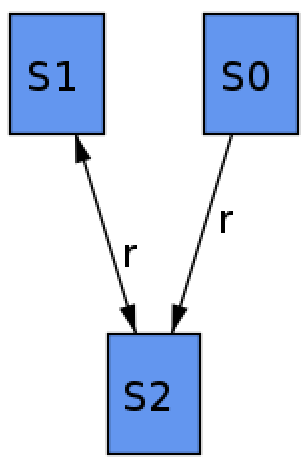}}
\end{minipage}
\\
(a) & (b) & (c) & (d)
\end{tabular}
\end{center}
\caption{\label{fig:irreflexive-instances}Four example instances
  generated by the Alloy analyzer.  Each instance represents an
  irreflexive relation.  \CodeIn{S0}, \CodeIn{S1}, and \CodeIn{S2} are
  the atoms in signature \CodeIn{S}.  The directed edges represent the
  tuples in relation \CodeIn{r}.}
\end{figure}

This section illustrates how we employ LLMs for writing Alloy
formulas.  We describe the basics of Alloy and the notation for
creating sketching
problems~\cite{WangETALABZ2018ASketch,SolarLazemaPhD2008} as we
introduce them.

The following Alloy code introduces a set \CodeIn{S} (\CodeIn{sig}), a
binary relation \CodeIn{r: S$\times$ S}, and a \Intro{predicate}
(\CodeIn{pred}) named \CodeIn{Irreflexive} that characterizes an
irreflexive relation:

\begin{CodeOut}
\begin{Verbatim}
\BLUE{sig} S \{ r: \BLUE{set} S \}

\BLUE{pred} Irreflexive \{
  \GREEN{-- No element in S is related to itself}
  \BLUE{all} s, t: S | s->t \BLUE{in} r \BLUE{implies} s != t
\}

\BLUE{run} Irreflexive
\end{Verbatim}
\end{CodeOut}

The binary relation \CodeIn{r} is declared as a \Intro{field} in the
signature (\CodeIn{sig}) declaration of set \CodeIn{S}.  The keyword
\CodeIn{set} in the field declaration defines \CodeIn{r} to be an
arbitrary binary relation; alternatively, the keywords \CodeIn{one}
and \CodeIn{lone} can be used to define a total function and a partial
function respectively.  Any text between `\CodeIn{-{}-}' (or
`\CodeIn{$//$}') and the end-of-line is a comment and ignored by the
Alloy analyzer.  The body of the predicate \CodeIn{Irreflexive} uses a
universally quantified (\CodeIn{all}) formula to define irreflexivity
as:
``$\forall s,t \in S ~|~ (s, t) \in r \Rightarrow s \neq t$''.
The Alloy operator `\CodeIn{->}' is Cartesian product.  The
\CodeIn{run} command defines a constraint solving problem and
instructs the Alloy analyzer to create an \Intro{instance}, i.e.,
valuation for the set \CodeIn{S} and relation \CodeIn{r} such that the
constraint specified in the predicate \CodeIn{Irreflexive} is
satisfied.  The analyzer solves each problem with respect to a
\Intro{scope}, i.e., a bound on the universe of discourse, which can
be explicitly stated in the \CodeIn{run} command.  The default bound
is 3, i.e., up to 3 atoms in each signature.  The analyzer supports
instance enumeration for satisfiable formulas.
Fig~\ref{fig:irreflexive-instances} shows 4~example instances created
by the Alloy analyzer.

\subsection{Natural language to Alloy}

Consider using an LLM to create an Alloy formula from a natural
language description of the desired property.  To illustrate, we run
the following query using ChatGPT:

\noindent\hfil\rule{0.99\textwidth}{1pt}\hfil

\vspace*{-2ex}
\begin{CodeOut}
\begin{verbatim}
Implement the following Alloy predicate Irreflexive as defined in the comments:
sig S { r: set S }
pred Irreflexive {
  -- No element in S is related to itself
}
Output only the formula in the predicate body.
\end{verbatim}
\end{CodeOut}

\vspace*{-3ex}

\noindent\hfil\rule{0.99\textwidth}{1pt}\hfil

\noindent{}ChatGPT produces the following output:

\noindent\hfil\rule{0.99\textwidth}{1pt}\hfil

\vspace*{-2ex}
\begin{CodeOut}
\begin{verbatim}
all s: S | s not in s.r
\end{verbatim}
\end{CodeOut}

\vspace*{-3ex}

\noindent\hfil\rule{0.99\textwidth}{1pt}\hfil

ChatGPT formulates the irreflexivity property using universal
quantification with just one variable.  The operator `\CodeIn{.}' in
Alloy is a relational composition.  Specifically, when \CodeIn{s} is a
scalar and \CodeIn{r} is a binary relation, the expression
`\CodeIn{s.r}' denotes the relational image of \CodeIn{s} under
\CodeIn{r}.  Thus, ChatGPT's formulation states that ``$\forall s\in S
~|~ s \not\in s.r$''.  Note how the formula structure of ChatGPT's
output differs quite a bit from how we wrote irreflexivity in the
predicate \CodeIn{Irreflexive}.

We validate ChatGPT's output using the Alloy analyzer by: (1)~creating
a new Alloy predicate \CodeIn{ChatGPTOutput} that contains the output
of ChatGPT; and (2)~checking that it is equivalent to our predicate
\CodeIn{Irreflexive} that serves as the ground truth:

\begin{CodeOut}
\begin{Verbatim}
\BLUE{pred} ChatGPTOutput \{
  \BLUE{all} s: S | s \BLUE{not} \BLUE{in} s.r
\}

\BLUE{check} \{ Irreflexive <=> ChatGPTOutput \}
\end{Verbatim}
\end{CodeOut}
The \CodeIn{check} command instructs the Alloy analyzer to find a
counterexample to the given formula, which asserts the logical
equivalence of our reference predicate and ChatGPT's output.  The
The alloy analyzer reports no counterexample and validates the correctness
of the output of ChatGPT (with respect to the default scope).

\subsection{Alloy to Alloy}
\vspace*{-1ex}

Next, consider using an LLM to create a new Alloy formula equivalent
to a given property in Alloy.  To illustrate, we run the following
query using DeepSeek:

\noindent\hfil\rule{0.99\textwidth}{1pt}\hfil

\vspace*{-2ex}
\begin{CodeOut}
\begin{verbatim}
Give an Alloy formula that is equivalent to the following predicate Irreflexive:
sig S { r: set S }
pred Irreflexive {
  all s, t: S | s->t in r implies s != t
}
Output only the formula.
\end{verbatim}
\end{CodeOut}

\vspace*{-3ex}

\noindent\hfil\rule{0.99\textwidth}{1pt}\hfil

\noindent{}DeepSeek produces the following output:

\noindent\hfil\rule{0.99\textwidth}{1pt}\hfil

\vspace*{-2ex}
\begin{CodeOut}
\begin{verbatim}
all s: S | s->s not in r
\end{verbatim}
\end{CodeOut}
\vspace*{-3ex}

\noindent\hfil\rule{0.99\textwidth}{1pt}\hfil

\noindent{}DeepSeek uses universal quantification with one variable
and states that ``$\forall s\in S ~|~ (s, s) \not\in r$''.  Note the
difference in the logical structure of DeepSeek's output and the input
formula we provided it.

We use the Alloy analyzer once again to validate the LLM's output.
The analyzer reports that no counterexample is found, i.e., the output
of DeepSeek is correct (with respect to the default scope).

\subsection{Sketching Alloy}
\vspace*{-1ex}

Next, consider using an LLM to complete the
sketch~\cite{SolarLazemaPhD2008,WangETALABZ2018ASketch} of an Alloy
model.  To illustrate, we run the following query using ChatGPT (in a
fresh session):

\noindent\hfil\rule{0.99\textwidth}{1pt}\hfil

\vspace*{-2ex}
\begin{CodeOut}
\begin{verbatim}
Complete the following sketch of the Alloy predicate Irreflexive with
respect to the property defined in the comments:
sig S { r: set S }
pred Irreflexive {
  -- No element in S is related to itself
  all s, t: S | s->t in r implies \E,e\ \O,o\ \E,e\
}
e := {| S|s|t|(s|t)->(s|t)|(s|t).r |}
o := {| in|!in|=|!= |}
Output only the completed predicate.
\end{verbatim}
\end{CodeOut}
\vspace*{-3ex}

\noindent\hfil\rule{0.99\textwidth}{1pt}\hfil

\noindent{}ChatGPT produces the following output:

\noindent\hfil\rule{0.99\textwidth}{1pt}\hfil

\vspace*{-2ex}
\begin{CodeOut}
\begin{verbatim}
pred Irreflexive {
  -- No element in S is related to itself
  all s, t: S | s -> t in r implies s != t
}
\end{verbatim}
\end{CodeOut}
\vspace*{-3ex}

\noindent\hfil\rule{0.99\textwidth}{1pt}\hfil

The sketch in the input query follows the notation used in previous
work~\cite{SolarLazemaPhD2008,WangETALABZ2018ASketch}.  There are
three \Intro{holes} in the sketch: two expression holes
(\CodeIn{$\backslash$E,e$\backslash$}) and one operator hole
(\CodeIn{$\backslash$O,o$\backslash$}).  The sketch defines a set of
possible completions for each hole.  Each expression hole can be
completed with one of the following nine~expressions: \CodeIn{\{`S',
  `s', `t', `s->s', `s->t', `t->s', `t->t', `s.r', `t.r'\}}.  The
operator hole can be completed with one of the following
four~operators: \CodeIn{\{`in', `!in', `=', `!='\}}.

Note that ChatGPT's output faithfully conforms to the structure of the
input sketch and each hole is completed appropriately.  Note also that
ChatGPT was given no directions regarding what a sketch is, let alone
how to complete it.  Nonetheless, the output is in fact identical to
our ground truth predicate, and is indeed correct.


\section{Methodology}
\label{sec:approach}

\begin{figure}[!t]
\centering
\includegraphics[width=0.5\textwidth]{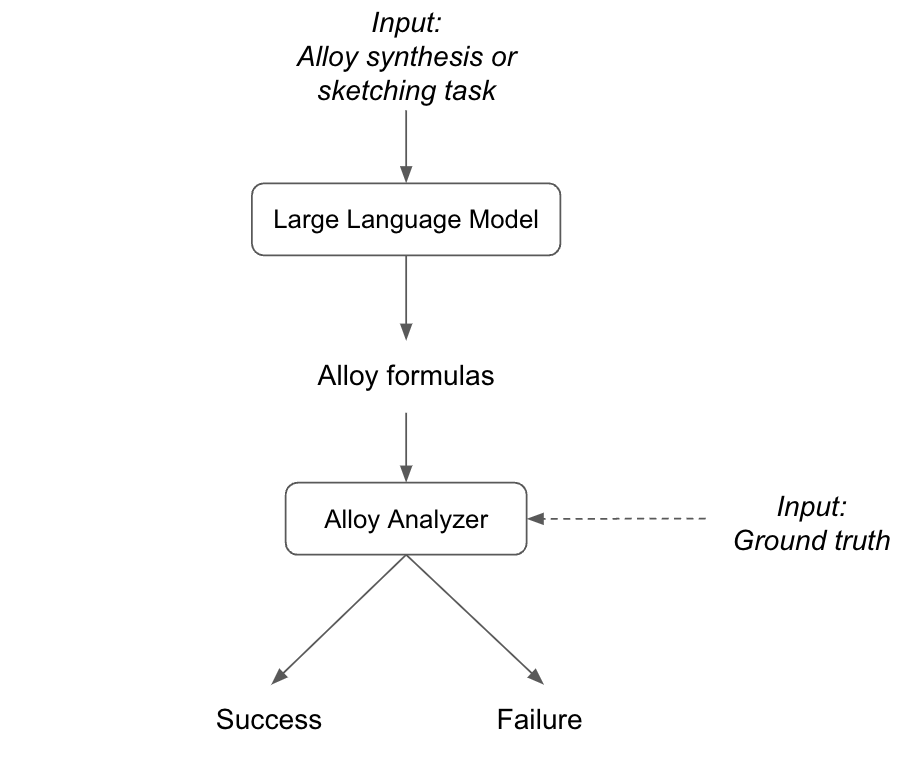}
\caption{Workflow. For each synthesis or sketching task, we create an input query for the LLM such that the query contains the target property in natural language or Alloy (depending on the kind of task), run the query, get the LLM's output, and use the Alloy analyzer to validate it with respect to a reference (ground truth) formula.}
\label{fig:workflow}
\end{figure}

We consider the following three methods for employing large language models (LLMs) to create Alloy formulas to investigate the capabilities and limitations of LLMs in writing Alloy:

\begin{enumerate}
\item
{\bf English to Alloy}. We employ LLMs to write complete Alloy formulas in multiple different ways from given natural language descriptions (in English);
\item
{\bf Alloy to Alloy}. We employ LLMs to create multiple alternative but equivalent formulas in Alloy with respect to given formulas in Alloy; and
\item
{\bf Sketch to Alloy}. We employ LLMs to complete sketches~\cite{SolarLazemaPhD2008,WangETALABZ2018ASketch} of Alloy
formulas and populate the holes in the sketches by synthesizing Alloy
expressions and operators so that the completed formulas accurately
represent the desired properties (that are given in natural language).  \end{enumerate}

\begin{table}[!t]
\begin{tabular}{r@{\hskip 0.2cm}|l|p{4cm}|p{5cm}}
& \multicolumn{1}{c|}{\Intro{Property}} & \multicolumn{1}{c|}{\Intro{Natural language desc.}} & \multicolumn{1}{c}{\Intro{Reference Alloy formula}}\\
\hline
1 & DAG & Directed acyclic graph &
\begin{lstlisting}[style=AlloyTable]
all n: Node | n !in n.^link
\end{lstlisting} \\
\hline
2 & Cycle & Graph with directed cycle &
\begin{lstlisting}[style=AlloyTable]
some n: Node | n in n.^link
\end{lstlisting} \\
\hline
3 & Circular & The number of nodes is equal to the number of edges and the graph has a directed cycle that visits all nodes &
\begin{lstlisting}[style=AlloyTable]
#Node = #link
all n: Node | one n.link
all m, n: Node | m in n.^link
\end{lstlisting} \\
\hline
4 & Connex & For every pair of elements in S, either the first is related to the second or vice versa &
\begin{lstlisting}[style=AlloyTable]
all s, t: S |
  s->t in r or t->s in r
\end{lstlisting} \\
\hline
5 & Reflexive & Every element in S is related to itself &
\begin{lstlisting}[style=AlloyTable]
all s: S | s->s in r
\end{lstlisting} \\
\hline
6 & Symmetric & If element x in S is related to y, then y is also related to x &
\begin{lstlisting}[style=AlloyTable]
all s, t: S |
  s->t in r implies t->s in r
\end{lstlisting} \\
\hline
7 & Transitive & If element x in S is related to y and y is related to z, then x is also related to z &
\begin{lstlisting}[style=AlloyTable]
all s, t, u: S |
  s->t in r and t->u in r
    implies s->u in r
\end{lstlisting} \\
\hline
8 & Antisymmetric & If element x in S is related to y and y is related to x, then x and y are the same element &
\begin{lstlisting}[style=AlloyTable]
all s, t: S |
  s->t in r and t->s in r
    implies s = t
\end{lstlisting} \\
\hline
9 & Irreflexive & No element in S is related to itself &
\begin{lstlisting}[style=AlloyTable]
all s, t: S |
  s->t in r implies s != t
\end{lstlisting} \\
\hline
10 & Functional & Every element in S is related to at most one element (making r a partial function) &
\begin{lstlisting}[style=AlloyTable]
all s: S | lone s.r
\end{lstlisting} \\
\hline
11 & Function & Every element in S is related to exactly one element (making r a total function) &
\begin{lstlisting}[style=AlloyTable]
all s: S | one s.r
\end{lstlisting} \\
\hline
\end{tabular}
\vspace*{2ex}
\caption{Subject properties. The table lists for each property, its
  natural language description that defines the corresponding natural
  language to Alloy task, and its reference formulation in Alloy that
  defines the corresponding Alloy to Alloy
  task.}\label{tab:subjects-synthesis}
\vspace*{-4ex}
\end{table}

\begin{table}[!h]
\centering
\begin{tabular}{p{12cm}}
\hline
\begin{lstlisting}[style=AlloyTable]
pred DAG {
  // Directed acyclic graph
  all n: Node | \E,e\ \CO,co\ \E,e\
}
co := {| =|in|!=|!in |}
e := {| Node|n|((Node|n).(*|^)link) |}
\end{lstlisting} \\ \hline

\begin{lstlisting}[style=AlloyTable]
pred Cycle {
  // Graph with directed cycle
  some n: Node | \E,e\ \CO,co\ \E,e\
}
co := {| =|in|!=|!in |}
e := {| Node|n|((Node|n).(*|^)link) |}
\end{lstlisting} \\ \hline

\begin{lstlisting}[style=AlloyTable]
pred Circular {
  // The number of nodes is equal to the number of edges and the graph has a directed cycle that visits all nodes
#Node = #link
  all n: Node | one n.link
  all m, n: Node | \E,e\ \CO,co\ \E,e\
}
co := {| =|in|!=|!in |}
e := {| (Node|m|n).(*|^)link |}
\end{lstlisting} \\ \hline

\end{tabular}
\vspace*{2ex}
\caption{Sketches for Alloy specifications for Properties 1--3.}
\vspace*{-8ex}
\label{tab:sketches-1-3}
\end{table}

Figure~\ref{fig:workflow} graphically illustrates our approach.
For each synthesis or sketching task, we create an input query for the LLM such that the query contains the target property in natural language or Alloy (depending on the kind of task), run the query, get the LLM's output, and run the Alloy analyzer to validate it with respect to a ground truth formula, which we provide to the analyzer. There are three possible outcomes of running the Alloy analyzer: (1) the LLM's answer is correct (when the analyzer does not find a counterexample to the equivalence of the LLM's answer and ground truth); (2) the LLM's answer has a syntax error (when the analyzer fails to compile the LLM's answer); and (3) the LLM's answer is wrong (when the analyzer finds a counterexample to the equivalence of the LLM's answer and ground truth). Note for "Alloy to Alloy" synthesis tasks, the ground truth formula is the reference formula given as input to the LLM. Note also that for any "English to Alloy" synthesis task and for any "Sketch to Alloy" sketching task, the input to the LLM does not include the ground truth formula.

We employ the LLMs directly as available for public use.  Specifically, we do not fine-tune them.  Moreover, the queries we write are minimalistic in their description of the problem domain and do not provide instructions to the LLM on how to approach solving any given task.

\subsection{Subject tasks}

We use \NumSubjects~well-known properties of graphs and binary relations to create \NumTotalTasks~tasks for the LLMs to answer.  Three of the properties (DAG, Cycle, and Circular) are regarding edge-labeled graphs, and the remaining eight properties (Connex, Reflexive, Symmetric, Transitive, Antisymmetric, Irreflexive, Functional, and Function) are regarding binary relations.  In Alloy, in general, we can use one signature $S$ and one binary relation $r: S\times S$ to represent either an edge-labeled graph or a binary relation. However, in view of the specific domain of graphs, we name the signature `\CodeIn{Node}' and the binary relation `\CodeIn{link}' when creating the tasks relating graph properties. For the tasks relating properties of binary relations, we name the signature `\CodeIn{S}' and the relation `\CodeIn{r}'.

For each property, we create 2~kinds of synthesis tasks: (1) create 20~unique Alloy formulas that represent the given natural language description of the property; and (2) create 20~unique Alloy formulas that are equivalent to the given Alloy formula that captures the property, which is also included as a natural language comment in the prompt.  In addition, for each property, we create one sketching task: complete the given sketch of the property with respect to its natural language description that is included as a comment in the prompt.  Thus, for each property, we have a total of 3~tasks for the LLM to answer.

Table~\ref{tab:subjects-synthesis} lists each property, its natural language description, and a reference (ground truth) formula that characterizes it in Alloy. Moreover, Tables~\ref{tab:sketches-1-3}, \ref{tab:sketches-4-8} (Appendix), and \ref{tab:sketches-9-11} (Appendix) list each property, its sketch that defines the corresponding sketching problem. Together these four tables summarize the key elements of our tasks for the LLMs. To illustrate, consider the DAG property.  Figure~\ref{fig:three-tasks-for-DAG} describes the actual prompts we run against each LLM for this property.

\begin{figure}[!p]
\centering
\begin{tcolorbox}[mytextbox]
Give me 20 unique solutions to the problem of synthesizing the body of the following Alloy predicate (without markdown or comments) with respect to the property described in the comments:
\begin{lstlisting}
sig Node {
  link: set Node
}
pred DAG{
  // Directed acyclic graph
  // your code go here
}
\end{lstlisting}
\end{tcolorbox}
(a) "English to Alloy" task\\
\begin{tcolorbox}[mytextbox]
Give me 20 unique solutions to the problem of synthesizing the body of the following Alloy predicate (without markdown or comments) with respect to the property described in the comments:
\begin{lstlisting}
sig Node {
  link: set Node
}
pred DAG{
  // Directed acyclic graph
  all n: Node | n !in n.^link
}
\end{lstlisting}
\end{tcolorbox}
(b) "Alloy to Alloy" task\\
\begin{tcolorbox}[mytextbox]
Complete the following sketch of the Alloy predicate (without markdown or comments) by selecting values for the holes with respect to the given constraints such that the predicate is correct with respect to the property described in the comments:

\begin{lstlisting}
sig Node {
  link: set Node
}
pred DAG {
  // Directed acyclic graph
  all n: Node | \E,e\ \CO,co\ \E,e\
}

co := {| =|in|!=|!in |}
e := {| Node|n|((Node|n).(*|^)link) |}
\end{lstlisting}
\end{tcolorbox}
(c) "Sketch to Alloy" task
\caption{Three tasks for the LLMs with respect to the DAG property.}
\label{fig:three-tasks-for-DAG}
\end{figure}

In a predicate sketch, certain components of the predicate are placeholder holes~\cite{WangETALABZ2018ASketch}. These holes can be of different forms, e.g., comparison operator holes, expression holes, and quantifier holes.  For all our sketching tasks, we only use two kinds of holes: comparison operator holes and expression holes. A predicate sketch includes a definition of the sets of possible values that each hole can be completed with.  These sets are typically defined using regular expressions~\cite{SolarLazemaPhD2008}.  For our DAG sketching task, the comparison operator hole may be completed with one of four possible values from the set \{ `\CodeIn{=}', `\CodeIn{in}', `\CodeIn{!=}', `\CodeIn{!in}'\}, and each expression hole may be completed with one of six possible values from the set \{ `\CodeIn{Node}', `\CodeIn{n}', `\CodeIn{Node.*link}', `\CodeIn{Node.\^{}link}', `\CodeIn{n.*link}', `\CodeIn{n.\^{}link}' \}.

\section{Evaluation}
\label{sec:evaluation}

We perform an experimental evaluation using two popular large language models, namely OpenAI o3-mini and DeepSeek R1. We answer the following research questions:

\textbf{RQ1 (English to Alloy)}: What is the performance of large language models in synthesizing multiple Alloy formulas that represent the input property that is given in natural language?

\textbf{RQ2 (Alloy to Alloy)}: What is the performance of large language models in synthesizing multiple Alloy formulas that are equivalent to the input Alloy formula that describes the desired property that is also given in natural language?

\textbf{RQ3 (Sketch to Alloy)}: What is the performance of large language models in completing the sketch of a property that is characterized by the input sketch and described in natural language?

\subsection{RQ1 (English to Alloy)}

\begin{table*}[!t]
\centering
\begin{tabular}{l|ccc|ccc}
\toprule
 & \multicolumn{3}{c|}{OpenAI o3-mini} & \multicolumn{3}{c}{DeepSeek R1} \\
Property & Correct & \hspace*{2ex}Syntax error\hspace*{2ex} & Wrong & Correct & \hspace*{2ex}Syntax error\hspace*{2ex} & Wrong \\ \hline
Antisymmetric & 16 & 4 & 0 & 19 & 1 & 0 \\
Circular      & 10 & 0 & 10 & 15 & 0 & 5 \\
Connex        & 8  & 10 & 2  & 14 & 4 & 2 \\
Cycle         & 5  & 4  & 11 & 8  & 2 & 10 \\
DAG           & 7  & 4  & 9  & 11 & 6 & 3 \\
Function      & 14 & 1  & 5  & 10 & 7 & 3 \\
Functional    & 15 & 5  & 0  & 15 & 3 & 2 \\
Irreflexive   & 10 & 10 & 0  & 15 & 4 & 1 \\
Reflexive     & 1  & 1  & 18 & 16 & 3 & 1 \\
Symmetric     & 6  & 14 & 0  & 16 & 4 & 0 \\
Transitive    & 15 & 4  & 1  & 19 & 1 & 0 \\
\bottomrule
\end{tabular}
\vspace*{2ex}
\caption{English to Alloy synthesis results.  For each property and
  each LLM, the table shows the number of correct formulas
  (\emph{Correct}), the number of syntactically invalid formulas
  (\emph{Syntax error}), and the number of syntactically valid but
  semantically wrong (\emph{Wrong}) formulas generated by the LLMs
  when asked to create 20 unique solutions.}
\label{tab:EnglishToAlloy-results}
\end{table*}

Table~\ref{tab:EnglishToAlloy-results} presents the results for each of the two LLMs. For each synthesis task, both LLMs find at least one valid solution, i.e., create an Alloy formula that is equivalent to the ground truth.

For OpenAI o3-mini, the number of correct formulas varies between 1 (for \CodeIn{Reflexive} property) and 16 (for \CodeIn{Antisymmetric} property); for 6 (out of 11) properties, it creates at least 10 correct formulas. For DeepSeek R1, the number of correct formulas varies between 8 (for \CodeIn{Cycle} property) and 19 (for \CodeIn{Antisymmetric} and \CodeIn{Transitive} properties); for 10 (out of 11) properties, it creates at least 10 correct formulas.

Overall, both LLMs perform well at creating Alloy formulas from their descriptions in natural language.





\subsection{RQ2 (Alloy to Alloy)}

\begin{table*}[!t]
\centering
\begin{tabular}{l|ccc|ccc}
\toprule
 & \multicolumn{3}{c|}{OpenAI o3-mini} & \multicolumn{3}{c}{DeepSeek R1} \\
Property & Correct & \hspace*{2ex}Syntax error\hspace*{2ex} & Wrong & Correct & \hspace*{2ex}Syntax error\hspace*{2ex} & Wrong \\ \hline
Antisymmetric & 18 & 1 & 1 & 9 & 8 & 3 \\
Circular      & 2 & 1 & 17 & 13 & 0 & 7 \\
Connex        & 15 & 3 & 2 & 16 & 3 & 1 \\
Cycle         & 13 & 2 & 5 & 16 & 2 & 2 \\
DAG           & 13 & 6 & 1 & 14 & 1 & 5 \\
Function      & 11 & 9 & 0 & 15 & 4 & 2 \\
Functional    & 8 & 6 & 6 & 7 & 8 & 5 \\
Irreflexive   & 16 & 4 & 0 & 15 & 0 & 5 \\
Reflexive     & 11 & 9 & 0 & 8 & 4 & 8 \\
Symmetric     & 10 & 9 & 1 & 18 & 1 & 1 \\
Transitive    & 17 & 3 & 0 & 17 & 3 & 0 \\
\bottomrule
\end{tabular}
\vspace*{2ex}
\caption{Alloy to Alloy synthesis results.  For each property and each
  LLM, the table shows the number of correct formulas, the number of syntactically invalid formulas, and the number of syntactically valid but
  semantically non-equivalent formulas generated by the
  LLMs when asked to create 20 unique solutions.}
\label{tab:AlloyToAlloy-results}
\vspace*{-2ex}
\end{table*}

Table~\ref{tab:AlloyToAlloy-results} presents the results for each of the two LLMs. For each synthesis task, both LLMs find at least two valid solutions, i.e., create an Alloy formula that is equivalent to the given input formula and its description in natural language.

For OpenAI o3-mini, the number of correct formulas varies between 2 (for \CodeIn{Circular} property) and 18 (for \CodeIn{Antisymmetric} property); for 9 (out of 11) properties, it creates at least 10 correct formulas. For DeepSeek R1, the number of correct formulas varies between 7 (for \CodeIn{Functional} property) and 18 (for \CodeIn{Symmetric} property); for 8 (out of 11) properties, it creates at least 10 correct formulas.

Overall, both LLMs perform well at creating Alloy formulas that are equivalent to given Alloy formulas and their descriptions in natural language.

\subsection{RQ3 (Sketch to Alloy)}

\begin{table*}[!t]
\centering
\begin{tabular}{l|c|c}
\hline
\Intro{Property} & \Intro{OpenAI o3-mini} & \Intro{DeepSeek R1}\\
\hline
Antisymmetric & \cmark & \cmark \\
\hline
Circular & \xmark & \cmark \\
\hline
Connex & \cmark & \cmark \\
\hline
Cycle & \cmark & \cmark \\
\hline
DAG & \cmark & \cmark \\
\hline
Function & \cmark & \xmark \\
\hline
Functional & \cmark & \cmark \\
\hline
Irreflexive & \cmark & \cmark \\
\hline
Reflexive & \cmark & \cmark \\
\hline
Symmetric & \cmark & \cmark \\
\hline
Transitive & \cmark & \cmark \\
\hline
\end{tabular}
\vspace*{2ex}
\caption{Alloy sketching results.  For each property and each LLM, the
  table shows whether the LLM successfully completes the sketching
  task.  For the property \Intro{Circular}, OpenAI o3-mini completed
  the input sketch with an Alloy formula that had a syntax error.
  Likewise, for the property \Intro{Function}, DeepSeek R1 completed
  the input sketch with an Alloy formula that had a syntax error.  For
  both these cases (where the LLM failed in its first attempt), we
  created a new query to inform it of the syntax-error and asked it to
  try again.  In each case the LLM succeeded to correctly complete the
  input sketch on the second attempt.}
\label{tab:sketching-results}
\vspace*{-2ex}
\end{table*}

Table~\ref{tab:sketching-results} presents the results for each of the two LLMs. Each LLM successfully completes all but 1 sketching task.

OpenAI o3-mini successfully completes sketches of all properties except \CodeIn{Circular}. The answer it gives for the sketching task for \CodeIn{Circular} has a syntax error.  We create a new query to inform OpenAI o3-mini of the syntax error and ask it to try again.  OpenAI o3-mini successfully completes the input sketch this time (second attempt).

DeepSeek R1 successfully completes sketches of all properties except \CodeIn{Function}. The answer it gives for the sketching task for \CodeIn{Function} has a syntax error.  We create a new query to inform DeepSeek R1 of the syntax error and ask it to try again.  DeepSeek R1 successfully completes the input sketch this time (second attempt).

Overall, both LLMs perform well at completing Alloy sketches.

\subsection{Discussion}

As the experimental results show, both the LLMs successfully create the desired Alloy formulas, which is indeed notable. We also note that the quality of solutions is high; an expert Alloy user can be expected to create solutions of such quality.  To illustrate, consider the following 11 equivalent formulas created by DeepSeek R1 as its answer to the "English to Alloy" task for the property \CodeIn{DAG}:

\begin{CodeOut}
\begin{verbatim}
 1. no ^link & iden
 2. all n: Node | n not in n.^link
 3. not some n: Node | n in n.^link
 4. all n: Node | no n.^link & n
 5. ^link = ^link - iden
 6. no iden & ^link
 7. all n: Node | (n.^link & n) = none
 8. all n: Node | #(n.^link & n) = 0
 9. all n: Node | lone (n.^link & n) => no (n.^link & n)
10. all n: Node | no n & n.^link
11. ^link in (^link - iden)
\end{verbatim}
\end{CodeOut}

Some of these formulas can be considered as relatively simple re-writes of each other, e.g., \CodeIn{\#}1 and \CodeIn{\#}7 that are equivalent because of the logical equivalence $a\wedge b~\equiv~b\wedge a$, or \CodeIn{\#}2 and \CodeIn{\#}3 that are equivalent because of the logical equivalence $\forall s\in S~|~\neg f(s)~\equiv~\neg\exists s\in S~|~f(s)$.  However, some others, e.g., ``\CodeIn{11. \^{}link in (\^{}link - iden)}'' (where \CodeIn{ident} is the built-in identity function), requires in-depth knowledge of Alloy to reason about its correctness.  Moreover, ``\CodeIn{9. all n: Node | lone (n.\^{}link \& n) => no (n.\^{}link \& n)}'' -- which states that for any node \CodeIn{n}, if the intersection of the relational image of \CodeIn{n} under the transitive closure of \CodeIn{link}, and the singleton set that contains just \CodeIn{n} has at most one element, then that intersection is empty -- is quite an unusual and rather interesting formulation of the DAG property.

It is also worth noting that the presence of the reference Alloy formula in the query (as in the "Alloy to Alloy" tasks) does not necessarily benefit the LLM in solving the synthesis problem for generating multiple equivalent solutions.  For example, DeepSeek R1 produces more correct formulas for 5 properties, fewer correct formulas for 5 properties, and an equal number of correct formulas for 1 property when synthesizing "English to Alloy" formulas than when synthesizing "Alloy to Alloy" formulas.  This is because enumerating equivalent but unique solutions requires a deeper level of understanding and having one solution available at the start does not necessarily help a lot.

\section{Related work}
\label{sec:related}

Large language models (LLMs) have enabled automation of many software
development and verification themes, including writing
code~\cite{BairiETALFSE2024,XiaETALFSE2025}, clarifying
requirements~\cite{MuETALFSE2024}, software
maintenance~\cite{DilharaETALFSE2024,JinETALFSE2024,JiangETALFSE2024,MaETALICSE2024,XuETALICSE2024,NamETALICSE2024},
software
testing~\cite{RyanETALFSE2024,LiuETALICSE2024,DinellaETALICSE2022},
debugging~\cite{WadhwaETALFSE2024,KangETAL2024}, constructing proofs
of theorems in automated provers~\cite{FirstETALFSE2024}, and
human-centric studies~\cite{WangETAL2024,ImranETALICSE2024}.  Our work
shares the spirit of previous work on using LLMs to formalize
specifications, e.g., postconditions~\cite{EndresETALFSE2024}, loop
invariants~\cite{ChakrabortyETALEMNLP2023}, or
Javadocs~\cite{JiangETALarXiv2024}.  In the specific context of Alloy,
LLMs were previously employed to repair faulty Alloy
models~\cite{Hasan2023,Alhanahnah2024}.  Indeed, repair and synthesis
are closely related; repair can be viewed as a restricted form of
synthesis where the faulty part of code that has been localized is
replaced with new code.  The topic of connecting LLMs and Alloy was
also briefly discussed on the Alloy community discourse
forum\footnote{\url{https://alloytools.discourse.group/t/the-use-of-alloy-and-llms/449}}.
To our knowledge, this paper presents the first study of LLMs in
solving the traditional synthesis and sketching problems for Alloy.

Program synthesis is a heavily studied
area~\cite{PueliRosnerICALP1989}.  A popular form of synthesis is
test-driven where user-provided tests are used to validate the
synthesized code.  SyPet~\cite{FengETAL2017} introduced the use of
Petri nets in creating sequences of method invocations for complex
APIs with respect to given tests.  EdSketch~\cite{HuaKhurshid2017} and
EdSynth~\cite{YangETAL2017} defined an optimized backtracking search
for completing Java sketches where test executions guided search
pruning.  Test-Driven Synthesis is defined an iterative method to build a
C$\#$ program such that it satisfies all the given tests~\cite{TDS}.
Component-based synthesis constructs programs by putting together
components from given libraries~\cite{oracleCBS}.

Program
sketching~\cite{Solar-LezamaETALCombSketchFinite2006,Solar-LezamaETALStencils2007,SolarLazemaPhD2008,jsketch,psketch,storyboardDS,AlurETAL2013,Singh2015},
pioneered by the Sketch system~\cite{SolarLazemaPhD2008}, offered an
exciting new advance in scaling program synthesis, where a partial
implementation is given and the goal is to complete
it~\cite{BodikJobstmannAlgoSynthesis2013,CodeHint,Feser2015,Kuncak:2010:CFS:1809028.1806632,Osera:2015:TPS:2813885.2738007,GveroETAL2011,MandelinETAL2005,FengETAL2017,synthesisRecursive,Singh2015}.
The Sketch system provided Java-like Sketch language for writing
partial programs, and deployed SAT and inductive synthesis in a
counterexample-guided loop to complete them.  JSketch enabled
sketching Java programs~\cite{jsketch} by translating Java to Sketch.
PSketch focused on concurrent data structures and enabled sketching
them~\cite{psketch}.

Researchers developed several approaches to assist Alloy users build
their models correctly, e.g., by highlighting UNSAT
cores~\cite{Shlyakhter2003,Torlak2008}, reducing
symmetries~\cite{TorlakJacksonTACAS2007,Shlyakhter2001,KhurshidETALSAT2003},
improving scenario
exploration~\cite{Alluminum,Nelson2017,SullivanETALICFEM2019},
supporting state
modeling~\cite{JacksonFeketeTACS01,FriasETALICSE05,JacksonVaziriISSTA2000,MarinovKhurshidASE2001,Sullivan2017EvaluatingSM},
and supporting testing a la traditional unit
testing~\cite{SullivanETALSPIN2014,SullivanETALICST2018,SullivanETALICST2017,WangETALICSE2018}.

ASketch~\cite{WangETALABZ2018ASketch} introduced sketching of Alloy
models in the spirit of the Sketch system.  The heart of ASketch is an
efficient enumerator for non-equivalent relational
expressions~\cite{WangETALABZ2018RexGen}.  We use the basic ASketch
approach for creating the sketching problems for the LLMs.  A key
difference is that ASketch, just like the Sketch system, requires test
cases to be given as input in order to complete the sketch and
validate the resulting code.  In contrast, our use of LLMs does not
require test cases; in addition to the sketch, we simply provide a
natural language description of the desired property to the LLMs.


\section{Conclusion}
\label{sec:conclusion}

This paper presented a three-fold use of large language models (LLMs)
in writing declarative formulas in the Alloy language.  One, LLMs were
employed to write complete Alloy formulas from given natural language
descriptions (in English).  Two, LLMs were employed to create
alternative but equivalent formulas in Alloy with respect to the given
Alloy formulas.  Three, LLMs were employed to complete sketches of
Alloy formulas and populate the holes in the sketches by synthesizing
Alloy expressions and operators so that the completed formulas
accurately represent the desired properties (given in natural
language).  An experimental evaluation using \NumSubjects{}
well-studied subject specifications and two popular LLMs, namely
ChatGPT and DeepSeek, were conducted.  A key aspect of the evaluation
was that for each synthesis problem (from English to Alloy and from
Alloy to Alloy), the LLMs were asked to generate multiple equivalent
but non-identical Alloy formulas as solutions, thus providing a deeper
look into how well the LLMs handle the semantic and syntactic
intricacies of the Alloy language.  The experimental results showed
that the LLMs generally performed quite well on synthesizing complete
Alloy formulas from input specifications given in natural language or
in Alloy, and were generally able to enumerate multiple unique
solutions.  Moreover, the LLMs were also successful at completing
given sketches of Alloy formulas.

LLMs hold much promise in enabling us to utilize the power of
specifications in building safe and reliable software systems.  Future
work will look at further utilizing LLMs in writing specifications,
and making LLMs an integral part of the process of writing
specifications -- just like they are today for writing
implementations.



\appendix

\newpage
\section*{Appendix: Alloy Sketches}
\label{app:sketches}
\vspace*{-2ex}

\begin{table}[!h]
\centering
\begin{tabular}{p{12cm}}
\hline
\begin{lstlisting}[style=AlloyTable]
pred Connex{
  // For every pair of elements in S, either the first is related to the second or vice versa
  all s, t: S | s->t in r or \E,e\ \CO,co\ \E,e\
}
co := {| =|in|!=|!in |}
e := {| r|s|t|((s|t)->(s|t)) |}
\end{lstlisting} \\ \hline

\begin{lstlisting}[style=AlloyTable]
pred Reflexive{
  // Every element in S is related to itself
  all s: S | \E,e\ \CO,co\ \E,e\
}
co := {| =|in|!=|!in |}
e := {| r|s|(s->s) |}
\end{lstlisting} \\ \hline

\begin{lstlisting}[style=AlloyTable]
pred Symmetric{
  // If element x in S is related to y, then y is also related to x
  all s, t: S | s->t in r implies \E,e\ \CO,co\ \E,e\
}
co := {| =|in|!=|!in |}
e := {| r|s|t|((s|t)->(s|t)) |}
\end{lstlisting} \\ \hline

\begin{lstlisting}[style=AlloyTable]
pred Transitive{
  // If element x in S is related to y and y is related to z, then x is also related to z
  all s, t, u: S | s->t in r and t->u in r implies \E,e\ \CO,co\ \E,e\
}
co := {| =|in|!=|!in |}
e := {| r|s|t|((s|t)->(s|t)) |}
\end{lstlisting} \\ \hline

\begin{lstlisting}[style=AlloyTable]
pred Antisymmetric{
  // If element x in S is related to y and y is related to x, then x and y are the same element
  all s, t: S | s->t in r and t->s in r implies \E,e\ \CO,co\ \E,e\
}
co := {| =|in|!=|!in |}
e := {| r|s|t|((s|t)->(s|t)) |}
\end{lstlisting} \\ \hline

\end{tabular}
\vspace*{2ex}
\caption{Sketches for Alloy specifications for Properties 4--8.}
\vspace*{-8ex}
\label{tab:sketches-4-8}
\end{table}

\begin{table}[!t]
\centering
\begin{tabular}{p{12cm}}
\hline

\begin{lstlisting}[style=AlloyTable]
pred Irreflexive{
  // No element in S is related to itself
  all s, t: S | s->t in r implies \E,e\ \CO,co\ \E,e\
}
co := {| =|in|!=|!in |}
e := {| r|s|t|((s|t)->(s|t)) |}
\end{lstlisting} \\ \hline

\begin{lstlisting}[style=AlloyTable]
pred Functional{
  // Every element in S is related to at most one element (making r a partial function)
  all s: S | \Q,q\ \E,e\
}
q := {| all|no|some|lone|one |}
e := {| r|s|(s.r) |}
\end{lstlisting} \\ \hline

\begin{lstlisting}[style=AlloyTable]
pred Function{
  // Every element in S is related to exactly one element (making r a total function)
  all s: S | \Q,q\ \E,e\
}
q := {| all|no|some|lone|one |}
e := {| r|s|(s.r) |}
\end{lstlisting} \\ \hline

\end{tabular}
\vspace*{2ex}
\caption{Sketches for Alloy specifications for Properties 9--11.}
\label{tab:sketches-9-11}
\end{table}

\bibliographystyle{splncs04}
\bibliography{main}

\end{document}